\documentclass[%
 reprint,
 floatfix,
superscriptaddress,
 amsmath,amssymb,
 aps
]{revtex4-2}
\usepackage{xcolor}
\usepackage{graphicx}
\usepackage{dcolumn}
\usepackage{bm}
\usepackage{subfig}
\usepackage{soul}
\usepackage{float}

\usepackage{ragged2e}
\DeclareCaptionJustification{justified}{\justifying}
\begin{document}

\preprint{APS/123-QED} 

\title{Blue-Detuned Magneto-Optical Trap of Molecules}

\author{Justin J. Burau}

\author{Parul Aggarwal}

\author{Kameron Mehling}

\author{Jun Ye}

\affiliation{
JILA, National Institute of Standards and Technology and the University of Colorado, Boulder, Colorado 80309-0440 \\ 
Department of Physics, University of Colorado, Boulder, Colorado 80309-0390, USA
}

\date{\today}

\begin{abstract}
Direct laser cooling of molecules has reached a phase space density exceeding 10$^{-6}$ in optical traps, but with rather small molecular numbers. To progress towards quantum degeneracy, a mechanism is needed that combines sub-Doppler cooling and magneto-optical trapping (MOT) to facilitate near unity transfer of ultracold molecules from the MOT to a conservative optical trap. Using the unique energy level structure of YO molecules, we demonstrate the first blue-detuned MOT for molecules that is optimized for both gray-molasses sub-Doppler cooling and relatively strong trapping forces. This first sub-Doppler molecular MOT provides an increase of phase-space density by two orders of magnitude over any previously reported molecular MOT.
\end{abstract}

\maketitle

Achieving a quantum degenerate gas of directly laser-cooled molecules remains an outstanding goal for realizing many applications within fundamental precision studies~\cite{ACME2018, Cairncross2017}, ultracold chemistry~\cite{Carr2009, Tobias2022}, and quantum simulations~\cite{ni2018,blackmore2018}.  In recent years, coherently associated bialkali molecules, KRb and NaK, have been cooled to quantum degeneracy \cite{Marco2019, Schindewolf2022, Pan2022}, leading to the observation of dipolar-mediated spin many-body dynamics~\cite{Li2022}. Direct laser cooling of diatomic molecules \cite{Hummon2013, Zhelyazkova2014,Hemmerling2016, Lim2018,Mcnally2020}, in the past decade, has given an impetus to the goal of making a quantum gas of laser cooled molecules that allows for a much broader range of molecules to be used for scientific applications. Important advances include demonstrating magneto-optical trapping (MOT) \cite{Barry2014, Anderegg2017,Williams2017,collopy2018,Vilas2022} and cooling to sub-Doppler temperatures \cite{Truppe2017, Cheuk2018, ding2020,Vilas2022} for several directly cooled molecular species.

For a molecular sample to achieve quantum degeneracy, its temperature needs to be cooled to a few tens of nanokelvin in a tightly confining trap. The temperatures currently achievable for laser-cooled molecules in optical dipole traps are on the order of tens of $\mu$K  \cite{Cheuk2018, Anderegg2019, Langin2021, Hallas2022}. The lowest temperature achieved is with YO molecules at 1 $\mu$K in an optical lattice \cite{WU2021}. One of the viable routes for achieving degeneracy is evaporative cooling of the molecules in a conservative trap~\cite{Valtolina2020, Li2021}. However, the densities in conservative traps of laser-cooled molecules are not quite high enough to initiate an efficient evaporation process. The primary reason for this is the low transfer efficiency of the molecules from magneto-optical traps into conservative traps. Typical red-detuned molecular MOTs are hot ($\sim$mK), producing large clouds with low densities. Sub-Doppler cooling via gray molasses (GMC), which follows the red-detuned MOTs, reduces the temperature of the molecules but the cloud sizes stay large in the absence of a confining force \cite{Cheuk2018, ding2020}. 

In this Letter, we report an effective solution to this problem by demonstrating the first blue-detuned MOT (BDM) of molecules. In our BDM for YO, we simultaneously achieve sub-Doppler cooling and a magneto-optical restoring force, with tunable operations between GMC and MOT. This signifies the first molecular MOT with temperatures well below the Doppler limit. With a corresponding compression in size,  the number density in the BDM is 50 times higher than that of a dual-frequency MOT (DFM). As a result, we achieve phase-space densities several orders of magnitude higher than any previous molecular MOT.

\begin{figure}[t!]
\includegraphics[width= 1 \linewidth]{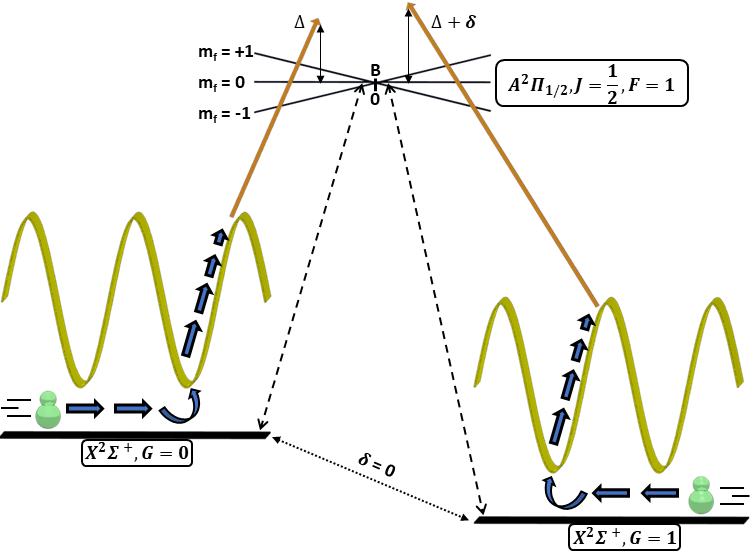}
\caption{
\label{fig:bdm_mechanism} The basic physical mechanism of the BDM. GMC relies on a motion-induced Sisyphus mechanism through a light-induced Stark shift gradient in the molecular ground-state of X$^{2}\Sigma^{+}$. While most of the molecules are cold and thus remain in the dark state, fast-moving molecules can undergo non-adiabatic transitions to a bright state. These molecules lose kinetic energy as they climb the Stark potential hill. Near the intensity maximum they will scatter a photon in a magnetic quadrupole field gradient which provides the MOT restoring force. Cooling is further enhanced in YO by coupling both ground-state manifolds in a $\Lambda$-type configuration as shown diagrammatically. By detuning the  two-photon resonance $\delta$ we can transition between robust coherent population trapping and MOT restoring force.
}
\end{figure}

The physical principles of the BDM are illustrated in Fig.~\ref{fig:bdm_mechanism}. The need for rotational closure to ensure continual optical cycling~\cite{Stuhl2008} calls for laser cooling and magneto-optical trapping of molecules to operate on type-II transitions, where the corresponding angular momenta change by $J$ $\rightarrow$ $J,J-1$. In Type-II transitions, the Doppler and sub-Doppler damping forces have opposite signs for a specific laser frequency detuning \cite{Jarvis2018}. Below a critical velocity on the order of $v_c$ $\approx$1 m/s, blue detuned laser light provides sub-Doppler cooling and Doppler heating \cite{Devlin2016}. This sub-Doppler cooling is known as GMC \cite{Weidemuller1994}, and it relies on non-adiabatic transitions between bright and dark states as the molecule moves through a light-induced Stark shift gradient \cite{Devlin2016,Weidemuller1994}. This cooling can be further enhanced by coupling multiple ground-state manifolds in a $\Lambda$-type configuration as shown in Fig.~\ref{fig:bdm_mechanism}, known as Raman-assisted-GMC or $\Lambda$-Enhanced cooling \cite{ding2020,Grier2013}. On two-photon resonance, a robust dark state is created by coherent population trapping \cite{Grier2013}. For the YO molecule, this cooling is extremely efficient and allows for $\mu$K temperatures over a large parameter space \cite{ding2020}.

\begin{figure}[H]
\includegraphics[width=1 \columnwidth]{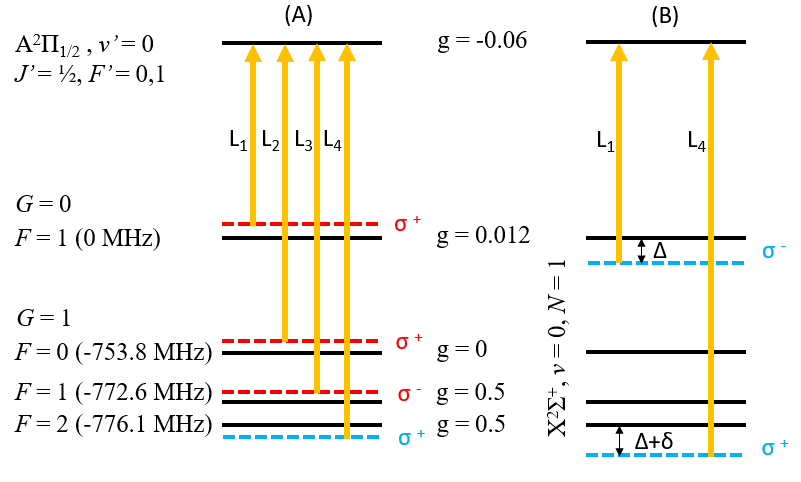}
\caption{\label{fig:level_structure} Relevant YO level structure and the transitions employed to address the hyperfine structure of $N =1$ rotational level of the main cooling laser for (A) Dual-Frequency MOT, (B) Blue-Detuned MOT.
}
\end{figure}

With the use of ground dark states, GMC is relatively insensitive to magnetic fields, when compared to typical sub-Doppler cooling on type-I transitions ($J$ $\rightarrow$ $J+1$)~\cite{ding2020, Devlin2016}. Based on this, a BDM was proposed and demonstrated in Rb atoms \cite{Jarvis2018}, by taking advantage of GMC cooling while providing a position-dependent scattering force \cite{Devlin2016}. The magnetic field sensitivity is further suppressed because of the unique hyperfine structure of YO with its small ground-state magnetic Lande $g$-factor. With a magnetic field up to $\approx$ 25~G, GMC remains effective for YO to produce a temperature more than three times below the Doppler-limit on two-photon resonance \cite{ding2020}. This means sub-Doppler cooling for YO should be particularly robust against typical magnetic quadrupole fields used in a MOT.   

Our experiment proceeds with a pulsed buffer gas beam source by ablation of a Y$_2$O$_3$ ceramic target. Chirped laser cooling is performed on the molecular beam  employing the X$^{2}\Sigma^{+}$ $(N=1)$ $\rightarrow$ A$^{2}\Pi_{1/2}$ $(J'=1/2)$ transition for rotational closure. After chirped laser cooling, molecules are loaded into a DFM, with details described in our previous work \cite{ding2020}. Figure~\ref{fig:level_structure}(a) shows the four laser beams that address the hyperfine structure in the $N = 1$ rotational level of the ground electronic state, X$^{2}\Sigma^{+}$. For a given combination of the excited-state $g$-factor sign, upper and lower angular momenta, quantization axis, and laser detuning, the different laser light polarizations are optimized to provide a strong trapping force \cite{Tarbutt2015}. Each of the four hyperfine components have equal intensity with the total intensity of all beams combined to be $I$ = 1.4$I_{sat}$ for a single pass, where $I_{sat}$ is defined for a two-level system. The magnetic field gradient is set to 12 G/cm. After the maximum DFM loading we apply a 2~ms pulse of GMC to rapidly cool the molecules from the DFM equilibrium temperature of 2 mK to $\sim$10 $\mu$K. We find that this pre-cooling increases the loaded fraction of molecules into the BDM by a factor of 2. This is accomplished by switching off the quadrupole field, and beams L$_{2}$ and L$_{3}$. The detuning of L$_{1}$ and L$_{4}$ is jumped to +3.13 $\Gamma/2\pi$, where $\Gamma$ is the linewidth of $A^{2}\Pi_{1/2}$. The detunings are defined relative to G = 0, F = 1 for beam L$_{1}$ and to G = 1, F = 2 for L$_{4}$, respectively. We keep the two-photon detuning on resonance, $\delta$ = 0. The intensity for L$_{1}$ and L$_{4}$ is increased to 10.9$I_{sat}$ each. 

After this 2~ms of GMC, L$_{1}$ is switched off and a new beam is turned on with opposite polarization as shown in Fig.~\ref{fig:level_structure}(b) to address the population in G = 0, F = 1. This is to give the right polarization for providing a trapping force, and is the critical step in transitioning from pure GMC in our previous work \cite{ding2020} to a BDM. Beam L$_{4}$ already has the correct detuning and polarization owing to the dual frequency mechanism \cite{ding2020}. The quadrupole field gradient is ramped back on to 4~G/cm over 40~ms and then held constant for the rest of the BDM. We use the same notation to describe the detunings as in Raman-assisted GMC: the one-photon detuning is defined as $\Delta$ and the two-photon detuning as $\delta$, depicted in Fig.~\ref{fig:level_structure}(b).

\begin{figure}[H]
\centering
\includegraphics[width=1.0 \columnwidth]{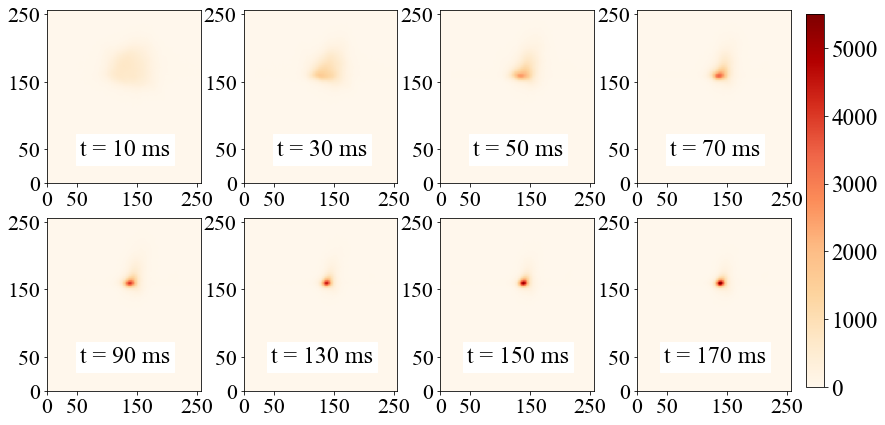}
\caption{
\label{fig:time_holds}
Evolution of the blue-detuned MOT as a function of different hold times (13~ms $in$-$situ$ exposure). The magnetic field is ramped from 0 to 4~G/cm over 40~ms and held constant for the remaining hold time, $\Delta$ = 0.44~$\Gamma$ and $\delta/2\pi$ = -264~kHz. Each image is 256 $\times$ 256 pixels, with a binsize 
 of 52~$\mu$m.
}
\end{figure}

Figure~\ref{fig:time_holds} shows the evolution of the YO molecular cloud in the BDM as a function of the MOT hold time $t$. Each image is taken $in$-$situ$ with a 13 ms exposure. At $t$ = 170~ms, with YO molecules having made $\sim$5 trap oscillations in the axial direction, the cloud reaches thermal equilibrium and is strongly compressed in size. The transfer efficiency of the molecules from DFM to BDM is about 50$\%$. 

We investigate the systematic dependence of the cloud temperature and size, thus the damping and trapping forces of BDM, on $I$, $\Delta$, and $\delta$. For small displacements from the trap center, the position-and velocity-dependent forces in the MOT can be described by  $m\ddot{r} = -\alpha \dot{r} -\kappa r$ where $\kappa$ is the spring constant, $\alpha$ is the damping coefficient, $m$ is the mass of YO, and $r$ is the displacement from the trap center. The spring constant $\kappa_{i,T}$ is obtained by using the equipartition theorem $\sigma^{2}_i$ $\kappa_{i,T}$  = $k_{B}T$, where $i$ = (axial, radial), $\sigma_{i}$ is the rms size, $k_B$ is Boltzmann's constant, and $T$ is the average temperature, defined as $T = T^{2/3}_{radial} \times T^{1/3}_{axial}$. We justify we are temperature limited in size as we measured no density-dependent lifetime in the BDM for optimized parameters. At temperature equilibrium we have a relationship between momentum diffusion and motional damping given by the Einstein relation in three dimensions.  

\begin{equation}
    \frac{D_p}{3\alpha} = k_BT .
        \label{eq:diff_and_alpha}
\end{equation}
$D_p$ is the momentum diffusion coefficient, defined as,
\begin{equation}
    D_p = \hbar^2k^2\Gamma_{scat},
    \label{eq:damping}
\end{equation}
\noindent where $\Gamma_{scat}$ is the scattering rate, $k$ is the wave-number, and $\hbar$ is the reduced Planck's constant. 

\begin{figure}[H]
\centering
\includegraphics[width=1 \columnwidth]{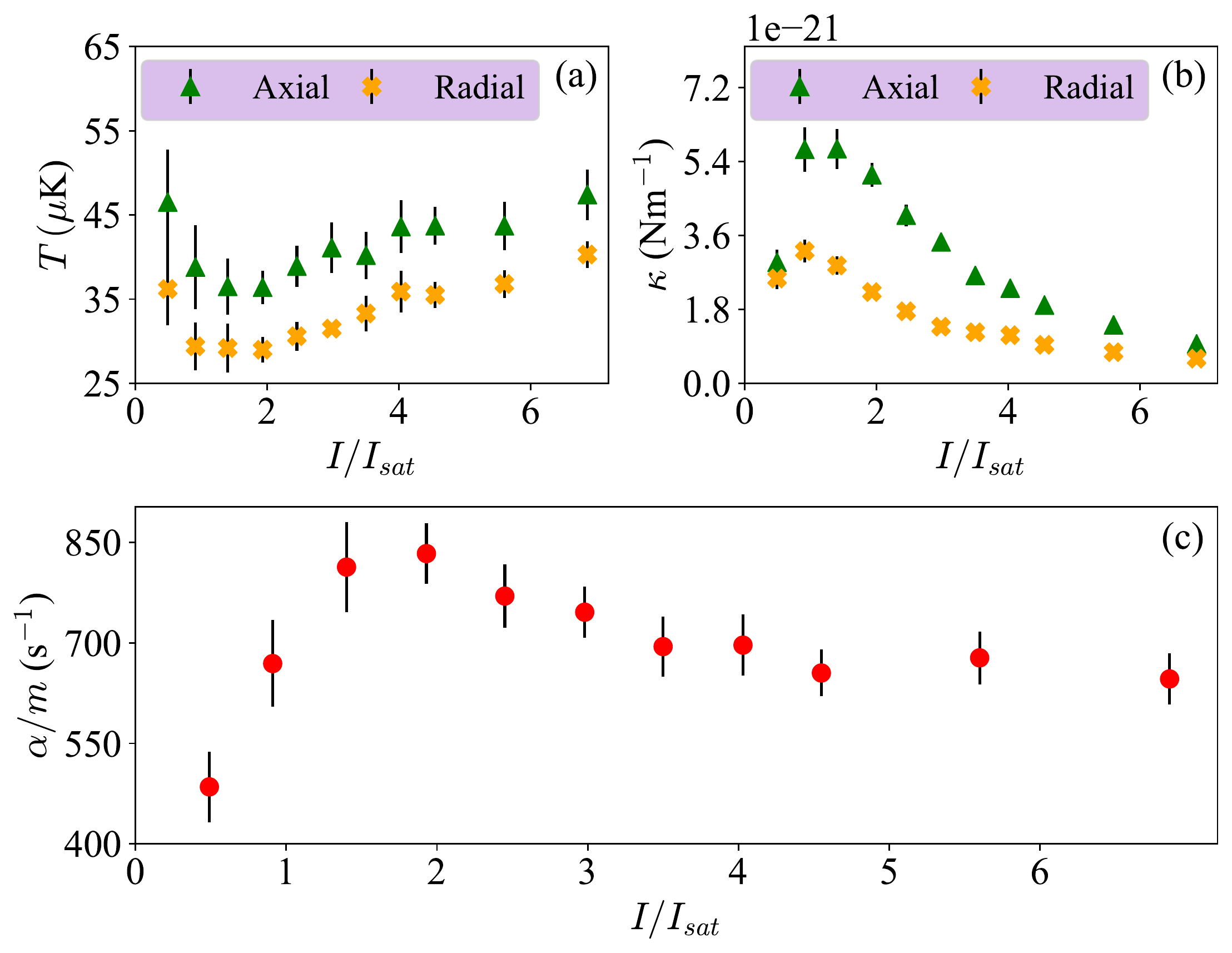}
\caption{
\label{fig:damping_constant_spring_constant}
Variation of the (a) temperature ($T$), (b) spring constant ($\kappa$), and (c) damping rate ($\alpha$/m) with intensity ($I$) of each sub-Doppler MOT hyperfine component, with the maximum magnetic field of 4 G/cm, $\Delta$ = 0.44 $\Gamma$ and $\delta/2\pi$ = -264~kHz.
}
\end{figure}

Figure~\ref{fig:damping_constant_spring_constant} details $T$, $\alpha$ and $\kappa$ of the BDM as a function of intensity $I$. Both L$_{1}$ and L$_{4}$ have equal intensity, indicated on the horizontal axis in terms of $I_{sat}$, and we measure $T$ along both the axial and radial directions using ballistic time-of-flight (TOF) expansion. The common one-photon detuning is $\Delta$ = 0.44 $\Gamma$ and the two-photon detuning is $\delta/2\pi$ = -264~kHz. The total hold time, $t$ is 100~ms. 

At relatively low intensities, $D_p$ increases linearly with $I$ and $\alpha$ should be relatively insensitive to $I$ in the regime where BDM is optimized \cite{Devlin2016}. Therefore, $T$ should increase linearly with $I$, as observed in Fig.~\ref{fig:damping_constant_spring_constant}(a). At lower intensities we see the temperature begin to rise and then increase rapidly. The cooling slows down for $I/I_{sat}$
 $<$~1 and the cloud may start falling under gravity, which leads to an increase in temperature that matches previous observations reported in \cite{Hodapp1995}. 

The dependence of $\kappa$ on $I$ is shown in Fig.~\ref{fig:damping_constant_spring_constant}(b).  The maximum spring constant value is observed under $I$ = 0.91 $I_{sat}$ at 5.7~$\times$~10$^{-21}$ N/m and 3.2~$\times$~10$^{-21}$ N/m in the axial and radial direction, respectively. This corresponds to a time period of the axial oscillation equal to 34~ms. This low trap oscillation frequency explains the slow compression displayed in Fig.~\ref{fig:time_holds}. The spring constant is observed to increase to a maximum value and then drops rapidly as $I$ increases. This behavior matches well with a full three-dimensional simulation of the spring constant based on Doppler theory \cite{Wallace1994}. We also note that the peak value of $\kappa$ in the axial direction is $\sim$2 times higher than that of the radial direction. This observation agrees with the expectation as the axial gradient is twice as strong as the radial gradient in the quadrupole field.    

To determine $\alpha$ we perform an additional light scattering rate measurement. We measure the scattering rate for the BDM by turning off $G = 1$ component of the $\nu$ = 1, 2 vibrational repumper. The BDM decay lifetime and the branching ratios to the states \cite{Zhang2020} for which the repumper is switched off are used to determine the scattering rate. The maximum scattering rate in the BDM is a factor of 3 lower than the DFM, which matches exactly the observation seen in \cite{Jarvis2018}. The relations expressed in Eqs.~\ref{eq:diff_and_alpha} and~\ref{eq:damping} then allow us to  determine $\alpha$ based on $T$ and $D_p$. Fig.~\ref{fig:damping_constant_spring_constant}(c) shows the damping rate ($\alpha/m$) as a function of $I$. The damping rate increases with $I$ to a peak value, then plateaus at higher values of intensity. This matches with what theory predicts as $\alpha$ should be intensity insensitive \cite{Devlin2016}.

\begin{figure}[htp]
\centering
\includegraphics[width=1 \columnwidth]{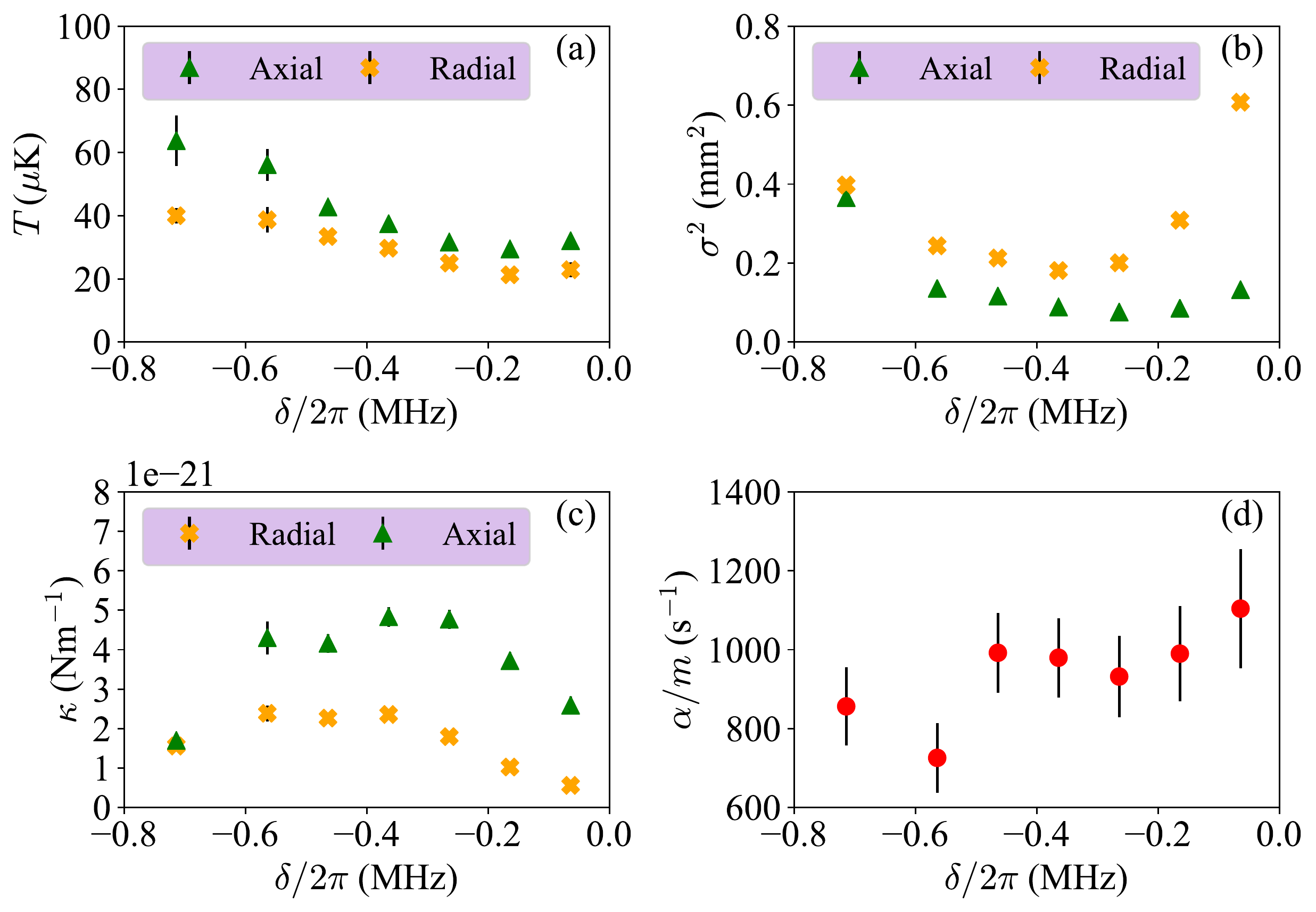}
\caption{
\label{fig:diffdetune_temp}
Variation of the (a) temperature ($T$), (b) quadrature size $(\sigma^{2}$), (c) spring constant ($\kappa$), and (d) damping rate ($\alpha/m$) in the blue MOT with differential detuning ($\delta$), with a maximum magnetic field gradient of 4~G/cm, $\Delta$ = 0.44 $\Gamma$ and I = 1.93~I$_{sat}$. We explicitly show the variation of $\sigma^{2}$ and $\kappa$ against $\delta$  to show the transition between BDM and GMC.
}
\end{figure}

We also investigated the BDM operating conditions as a function of $\delta$. Figures~\ref{fig:diffdetune_temp}(a) and \ref{fig:diffdetune_temp}(b) plot the measured $T$ and $\sigma^2$  respectively, under various values of $\delta$. The horizontal axis denotes the detuning from two-photon resonance between G = 0, F = 1 and G = 1, F = 2. The temperature plateaus and reaches a minimum value near $\delta$ = 0.  The quadrature size $\sigma^2$ is determined from extrapolating TOF to zero expansion time. We see a very prominent increase in $\sigma^2$ near $\delta$ = 0, along with the development of a large disparity in size between the radial and axial directions. As we detune away from resonance we reach a minimum value in $\sigma^2$, and then it increases again as the detuning increases, but the size in the axial and radial directions do not develop a large difference. In Fig.~\ref{fig:diffdetune_temp}(c) we plot $\kappa$ using the same method as before, and we see a corresponding trend to the size. On the other hand, the damping rate increases as we approach two-photon resonance, as shown in Fig.~\ref{fig:diffdetune_temp}(d). 

These observations can be explained as a transition of operational regimes between GMC and BDM as a function of $\delta$. For GMC to facilitate lowest temperatures, the excited population should be as low as possible for coherent population trapping. However, for BDM the excited-state population cannot be zero as there will be no position-dependent scattering force. As such, there is a fine balance of parameters that leads to efficient sub-Doppler cooling, along with an appreciable excited-state population to provide the MOT restoring force. By tuning to $\delta$ = 0, a coherent dark state is formed between the ground-state manifolds and the excited-state population drops rapidly. This explains the observation of the spring constant dropping towards zero and the size remaining large. We also see that at resonance, the damping rate approaches the highest value and the lowest temperatures are obtained. Thus, by tuning the two-photon detuning, we can easily transition from a regime where the molecules are in a 3D gray molasses to a BDM configuration where an appreciable trapping force appears. This is a unique feature of the BDM that is not found in typical Type-I transition optical molasses used for cooling most atomic species. It alludes to the fact that Type-II  transitions provide a close physical connection between GMC and BDM.  

\begin{figure}[h!]
\centering
\includegraphics[width=1 \columnwidth]{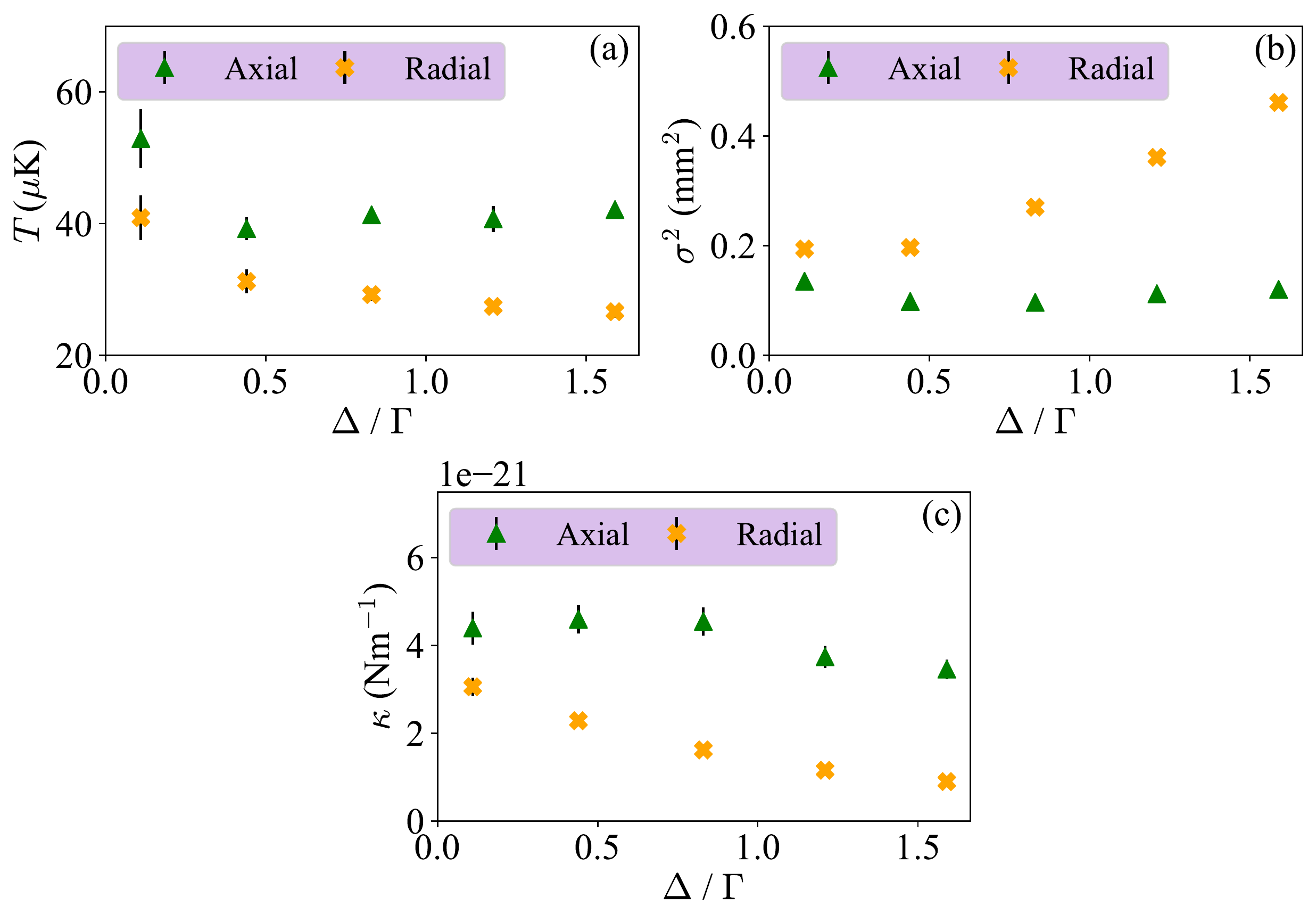}
\caption{
\label{fig:damping_constant_spring_constant_one_photon}
Variation of the (a) temperature ($T$), (b) quadrature size ($\sigma^{2}$), and (c) spring constant ($\kappa$) with one-photon detuning ($\Delta$) of each sub-Doppler MOT hyperfine component, with the maximum magnetic field of 4 G/cm, $\delta/2\pi$ = -264 kHz and I = 1.93~I$_{sat}$.
}
\end{figure}

In Fig.~\ref{fig:damping_constant_spring_constant_one_photon}, we optimize the temperature and size in the BDM against one-photon detuning $\Delta$. The variation of $T$ as a function of $\Delta$ is shown in Fig.~\ref{fig:damping_constant_spring_constant_one_photon}(a). The temperature stays roughly constant for all values of  $\Delta$ larger than  0.44 $\Gamma$. For smaller values of $\Delta$ there is an appreciable excited-state population which increases the temperature due to photon scattering. This trend can also be seen in the dependence of size and spring constant in Figs.~\ref{fig:damping_constant_spring_constant_one_photon}(b) and \ref{fig:damping_constant_spring_constant_one_photon}(c). The population in the excited-state at smaller $\Delta$ provides a spring constant that helps to reduce the size.  We find the optimal value of $\Delta$ to be 0.44 $\Gamma$ as it provides the best compression for a relatively low temperature.

The peak number density we have achieved so far is 2.5 $\times$ 10$^{8}$ cm$^{-3}$. The size of the cloud in the radial and axial direction is 225~$\mu$m and 202~$\mu$m respectively at a field gradient of 14 G/cm. This leads to a phase-space density in the BDM of 5 $\times$ 10$^{-9}$ with an average temperature of $T$ = 38(4) $\mu$K.

Further, applying a 1 ms pulse of free space GMC with the same parameters as described earlier after the BDM reduces the temperature to $T$ = 8.8(0.4) $\mu$K. With a nominal 4(1)$\%$ increase in the cloud size in both radial and axial directions. With this temperature and size we achieve a free-space phase-space density of 4.5 $\times$ 10$^{-8}$ which surpasses the highest free-space phase-space density even at this elevated temperature \cite{ding2020}. This denser molecular sample will be loaded into a crossed dipole trap, where we aim to achieve a near unity transfer. With an estimated trapping frequency of several hundred Hz in the radial direction \cite{WU2021}, and GMC cooling remaining effective inside a dipole trap, we anticipate a phase-space density well above 1 $\times$ 10$^{-6}$, together with a large number of YO molecules. Recently, inelastic molecular collision loss has been suppressed by electric field-induced resonance~\cite{Matsuda2020, Li2021} or microwave shielding \cite{Anderegg2021, Schindewolf2022}. Applying these techniques to YO evaporative cooling may open a path towards a molecular Bose-Einstein condensate. 

We thank L.Caldwell and E. B. Norrgard for careful reading of the manuscript and their useful comments. Funding support for this work is provided by ARO MURI, AFOSR MURI, NIST, and NSF PHY-1734006. 




\nocite{*}

\bibliography{main.bib}

\end{document}